\documentclass[
,final            
  ,                 
  ]
  {aipproc}
\layoutstyle{8x11double}

\begin{document}

\title[]{Measurement of identified charged hadron spectra in proton-proton collisions using the Inner Tracking System of the ALICE experiment at the LHC}


\author{Emanuele Biolcati \\ \small{on behalf of the ALICE collaboration}}{
  address={biolcati@to.infn.it}
}

\begin{abstract}
The measurement of the identified charged
hadron $p_{t}$ spectra using the ITS energy loss signal in the $pp$ data at $\sqrt{s}=900$~GeV collected by 
the ALICE experiment at LHC will be discussed.
It is performed using the Inner Tracking System (ITS) in stand-alone mode, both for track reconstruction  
and particle identification, allowing one to detect low momentum particles with $p_{t}$ below 200~MeV/c. A second method 
using the tracks reconstructed by both the ITS and the Time Projection Chamber (TPC) has also
been developed. The obtained results will be compared with the ones obtained from the TPC and from the Time Of 
Flight detector and used to extract the mean $p_t$ value.
\end{abstract}

\maketitle


\section{Introduction}
The measurement of identified charged hadron spectra in proton-proton collisions at~$\sqrt{s}$~=~900~GeV 
has been performed by the ALICE experiment on the data collected in~2009 at LHC.
The results provide basic insights into particle production processes and give the
possibility to test and tune commonly used event generators at LHC energies. In addition, when repeated
at higher energy, they will be the baseline for measurements in heavy-ion collisions.

The ALICE experiment features detectors that have multiple particle identification systems, including the Time
Projection Chamber (TPC), the Inner Tracking System (ITS), a dedicated time-of-flight system
(TOF). They allows ALICE to 
identify charged hadrons in the momentum range from~90~MeV/c up to~2.5~GeV/c. More details
about the ALICE design can be found in~\cite{alicejinst}. Further information about the identified particle
studies with this experiment can be found in~\cite{panosmadrid}.

The Inner Tracking System, close to the interaction point, is built of three sub-detectors, 
each of those contains two concentric Silicon layers. The
innermost part is the pixel detector (SPD), the middle part is the drift detector (SPD) and the 
outermost one is the strip detector (SSD). The ITS provides high
resolution tracking for the identification of secondary vertices of heavy flavour weak decays and
improved momentum resolution at moderate $p_{t}$. The ITS is also used for stand-alone tracking
to measure tracks at very low $p_{t}$ not reaching the TPC.

\section{ITS analysis}
\begin{figure}[h]
  \includegraphics[width=0.45\textwidth,height=4.5cm]{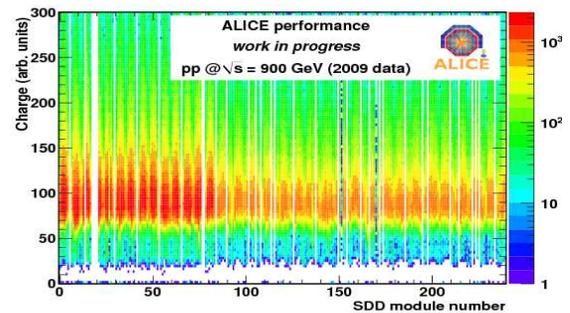}
  \caption{SDD charge distribution versus module number. More statistic for the inner layer (first 84 modules).}
	\label{fig0}
  \end{figure}

\begin{figure}
  \includegraphics[width=0.45\textwidth,height=4.5cm]{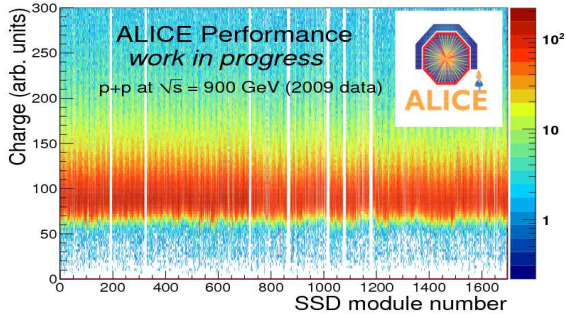}
  \caption{SSD charge distribution versus module number.}
	\label{fig00}
\end{figure}

The calculation of the identified charged hadron spectra in the ITS is based on the measurement of the particle energy
loss in 4 (SSD and SDD) out of 6 Silicon layers. The ITS sub-systems have been aligned and commissioned using the cosmic data 
taken by the ALICE experiment in 2008, for more details see \cite{dainese,mario,panos}.   
To achieve a precise charge calibration, the distribution of charge per unit path length has 
been measured and fine tuned for each module of the SDD and the SSD. In figures~\ref{fig0} and~\ref{fig00} it is possible 
to see that the peaks of the charge distributions are centered at the same value with a tolerance less than 5\%.


\begin{figure}[h]
  \includegraphics[width=0.45\textwidth]{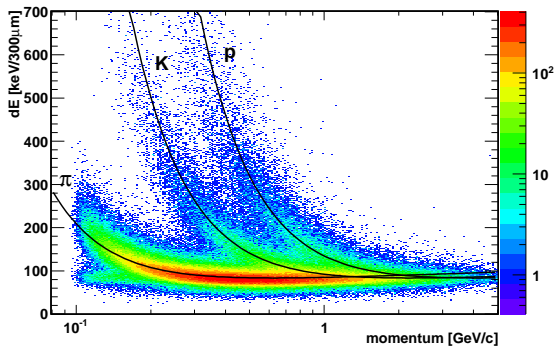}
  \caption{Energy loss signal versus momentum for the ITS stand-alone tracks. Three lines show the pion, kaon and proton signals 
	expected by the Bethe-Bloch formula.}
	\label{fig1}
\end{figure}

Two independent analyses have been performed using two different ITS track samples: the stand-alone ones, 
reconstructed using only information from the ITS (6 silicon layers) and the global ones, 
that use all the ALICE tracking detectors. In any case, only tracks with 3 or 4~points
localized in the SDD and SSD are used in the analysis. The
charge deposit measured in each layer is normalised to the energy loss per~300~$\mu$m (thickness
of the ITS layers) of Silicon using the track local angles as reconstructed by the
tracking algorithm. In the case of SDD clusters, a
linear correction for the dependence of the reconstructed
raw charge as a function on drift time due to the combined
effect of charge diffusion and zero suppression is
also applied \cite{mio}. For each track, ${\rm d}E/{\rm d}x$ is calculated using
a truncated mean: the average of the lowest two points
in case 4 points are measured, or a weighted sum of
the lowest (weight 1) and the second lowest point (weight
1/2) in case 3 points are measured.
Figure~\ref{fig1} shows the energy loss signal so obtained in the ITS 
as function of the momentum for stand-alone tracks, compared to the parametrised response
for pions, kaons and protons. This parametrisation in based on the Bethe-Bloch formula from~\cite{phobos}, 
	with an additional correction for the low momentum part.

\begin{figure}[h]
  \includegraphics[width=0.45\textwidth]{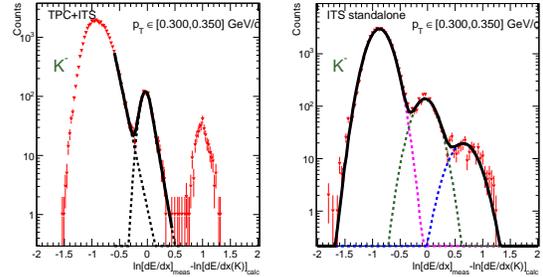}
  \caption{Distributions of the variable $\xi$ in a $p_t$ bin, in the kaon hypothesis mass, for the two ITS track samples. The fit curves are superimposed.}
	\label{fig2}
\end{figure}

For each track fulfilling the track quality requirements, the rapidity is calculated using three
different mass hypotheses: pion, kaon and proton. Tracks with $|y|<0.5$ are selected. For
each hypothesis in $p_t$ bins, a histogram is filled with the difference between the logarithm of the
measured energy loss signal in the ITS and expected value calculated using the variable 
$$
 \xi = \log[{\rm d}E/{\rm d}x]_{\rm mean} - \log[{\rm d}E/{\rm d}x]_{\rm calc}.
$$
This procedure gives a series of three histograms for each charged
particle in all $p_t$ bins. Figure~\ref{fig2} shows an example of these distributions for the kaon mass
hypothesis. 

\begin{figure}[h]
  \includegraphics[width=0.45\textwidth]{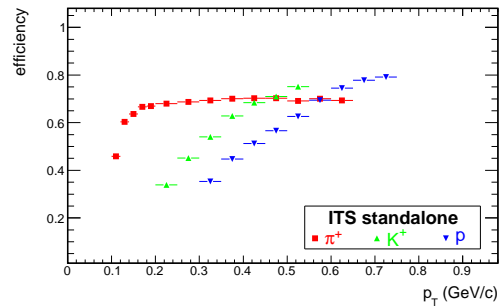}
  \caption{Overall efficiency of the ITS stand-alone versus $p_t$. Only positive pion, kaon and proton points plotted.}
	\label{fig3}
\end{figure}

The area below the peak centered at zero
is the raw yield of particles from the mass hypothesis in given $p_t$ bin. The raw yield is determined
by fitting the histograms with functions show in the figure~\ref{fig2}: a sum of three Gaussians
for the ITS stand-alone analysis and a sum of a Gaussian with an exponential tail
and the tail of the peak at lower ${\rm d}E/{\rm d}x$ for the TPC+ITS analysis. The raw yield is then
corrected for the contamination from secondary particles and the overall efficiency. This has been calculated by using
a Monte-Carlo simulation of the collisions and the 
detector in the same configuration of the data taking period. 
Figure~\ref{fig3} shows the overall efficiencies for the ITS stand-alone analysis for positive pions, kaons and protons. 
They take into account the geometric acceptance, the tracking efficiency, the dead channels and the K~decays. 

\section{Results}
\begin{figure}[h]
  \includegraphics[width=0.45\textwidth,height=5cm]{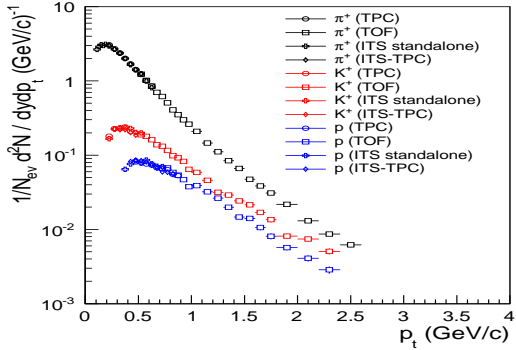}
  \caption{Identified particle spectra obtained by the four independent analyses using ITS, TPC and TOF (positive only).}
	\label{fig4}
\end{figure}

After the correction, performed separately for each analysis, the yields have been normalized to the number of the inelastic collisions. The first important result is the coherence between the four independent analyses: ITS stand-alone, ITS+TPC, TPC and TOF. In figure~\ref{fig4} the matching of the points for the three positive species is observable. It is thus possible to cover the whole transverse momentum range from~90~MeV/c to~2.5~GeV/c.

The spectra have been combined weighting for the uncorrelated systematic errors and fitted using the L\'evy function~\cite{abelev}: 
	$$
  \label{eq:1}
  \frac{{\rm d}^2N}{{\rm d}p_t{\rm d}y} = p_t \frac{{\rm d}N}
{{\rm d}y} \frac{(n-1)(n-2)}{nC(nC + m (n-2))} \left( 1 + \frac{m_t - m_0}{nC} \right)^{-n}.
 $$

\begin{figure}[h]
  \includegraphics[width=0.45\textwidth,height=5cm]{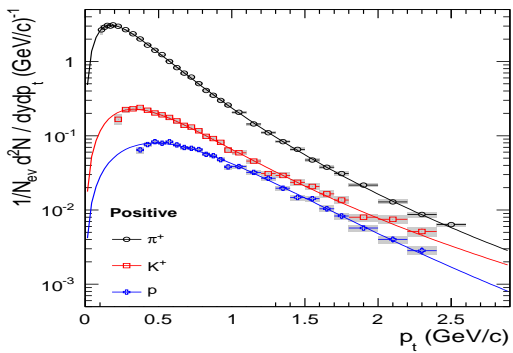}
  \caption{Transverse momentum spectra of positive hadrons from pp collisions at $\sqrt{s}$~=~900~GeV. Grey
	band: total $p_t$-dependent error (systematic plus statistical);
	normalization systematic error not plotted.}
	\label{fig5}
\end{figure}

In figure~\ref{fig5} the fit result for the positive particles is shown. 

\begin{figure}[h]
  \includegraphics[width=0.45\textwidth,height=6cm]{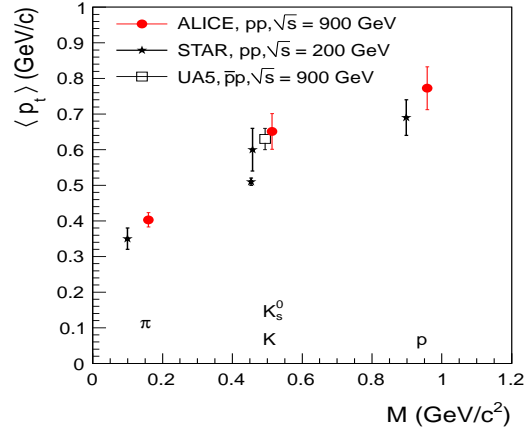}
  \caption{Mean $p_t$ versus the mass of the emitted
	particle obtained in pp collisions at $\sqrt{s}$~=~200~GeV
	and at 900~GeV and also in $\bar{p}p$ at $\sqrt{s}$~=~900~GeV.}
	\label{fig6}
\end{figure}

The mean transverse momentum, extracted from the fit, has been compared with the results 
obtained in $pp$ collisions at~$\sqrt{s}$~=~200~GeV and in
$\bar{p}p$ reactions at~$\sqrt{s}$~=~900~GeV. The mean $p_t$
rises very little with increasing $\sqrt{s}$ despite the fact that
the spectral shape clearly shows an increasing contribution from
hard processes. It was already observed that at RHIC energies the
increase $\langle p_t\rangle$, compared to studies at~$\sqrt{s}$~=~25~GeV, is
small (figure~\ref{fig6}).

\section{Conclusions}
This measurement is an important confirmation about the good operation of the central detectors of the ALICE experiment for $pp$ collisions. As a matter of fact, the analyses performed using different detectors and different approaches give a coherent result, allowing to cover a large transverse momentum range. 

The first identified hadron $p_t$ spectra from proton-proton collisions at LHC have been analyzed. In this way, the whole analysis framework has been tuned in order to repeat the study for $pp$ data at higher energies.


\bibliographystyle{aipproc}

\end{document}